\RequirePackage{fix-cm}
\documentclass[smallextended]{svjour3}       
\smartqed  
\usepackage{tablefootnote}
\usepackage{authblk}
\usepackage{amsmath, amsfonts, amssymb, color, graphicx, mathtools, enumerate, subfigure}
\usepackage{multirow, float, changepage}
\usepackage{natbib}
\usepackage{bm}
\usepackage{wrapfig}
\usepackage{color}
\usepackage{changes}
\usepackage{algorithm} 
\usepackage{algorithmic}

\usepackage[CJKbookmarks=true,
            bookmarksnumbered=true,
  bookmarksopen=true,
  colorlinks=true,
  citecolor=blue,
  linkcolor=blue,
  anchorcolor=blue,
  urlcolor=blue]{hyperref}
  
\usepackage[letterpaper, left=1.2truein, right=1.2truein, top = 1.2truein, bottom = 1.2truein]{geometry}
\usepackage{lipsum}
\usepackage{filecontents}

\usepackage{caption}

\newcommand{\D}{\mathcal{D}}
\newcommand{\G}{\mathcal{G}}
\newcommand{\EE}{\mathbb{E}}

\newcommand{\RR}{\mathbb{R}}

\newcommand{\sig}{{\sf sigmoid}}
\newcommand{\relu}{{\sf ReLU}}
\newcommand{\ramp}{{\sf ramp}}
\newcommand{\Lap}{\mathrm{Laplace}}
\newtheorem{thm}{\bf Theorem}

\newcommand{\HLL}[1]{\textcolor{black}{#1}}

\providecommand{\keywords}[1]
{
  {\small	
  \textbf{{Keywords:}} #1
  }
}
\usepackage{prettyref,soul}
\usepackage{varwidth}

\begin{document}

\title{Generative Adversarial Networks 
for Robust Cryo-EM Image Denoising\thanks{This research made use of the computing resources of the X-GPU cluster supported by the Hong Kong Research Grant Council Collaborative Research Fund: C6021-19EF. The research of Hanlin Gu and Yuan Yao is
supported in part by HKRGC 16308321, ITF UIM/390, as well as awards from Tencent AI Lab and Si Family
Foundation. We would like to thank Dr. Xuhui Huang for helpful discussions.}
}


\author{Hanlin Gu         \and
        Yin Xian \and
        Ilona Christy Unarta \and
        Yuan Yao 
}


\institute{Hanlin Gu \at
              Hong Kong University of Science and Technology, Hong Kong SAR China. \\
              \email{hguaf@connect.ust.hk}
		\and
		Yin Xian \at
              Hong Kong Applied Science and Technology Research Institute (ASTRI), Hong Kong SAR China. \\
              \email{poline3939@gmail.com}           
           \and
           Ilona Christy Unarta \at
           Hong Kong University of Science and Technology, Hong Kong SAR China. \\
           \email{icunarta@connect.ust.hk}           
           \and
           Yuan Yao \at
           Hong Kong University of Science and Technology, Hong Kong SAR China. \\
           \email{yuany@ust.hk}          
}

\date{Received: date / Accepted: date}

\maketitle

\tableofcontents

\vspace{15pt}


\begin{abstract}
 The Cryo-Electron Microscopy (Cryo-EM) becomes popular for macromolecular structure determination. However, \HLL{the 2D images captured by Cryo-EM} are of high noise and often mixed with multiple heterogeneous conformations and contamination, imposing a challenge for denoising. \HLL{Traditional image denoising methods and simple Denoising Autoencoder cannot work well when the Signal-to-Noise Ratio (SNR) of images is meager and contamination distribution is complex.} Thus it is desired to develop new effective denoising techniques to facilitate further research such as 3D reconstruction, 2D conformation classification, and so on. \HLL{In this chapter}, we approach the robust denoising problem for Cryo-EM images by introducing a family of Generative Adversarial Networks (GAN), called $\beta$-GAN, which is able to achieve \HLL{robust estimation of} certain distributional parameters under Huber contamination model with statistical optimality. \HLL{To address the denoising challenges, for example,} the traditional image generative model might be contaminated by a small portion of unknown outliers, $\beta$-GANs are exploited to enhance the robustness of Denoising Autoencoder. \HLL{Our proposed method} is evaluated by both a simulated dataset on the Thermus aquaticus RNA Polymerase (RNAP) and a real world dataset on the Plasmodium falciparum 80S ribosome dataset (EMPIAR-10028), in terms of Mean Square Error (MSE), Peak Signal-to-Noise Ratio (PSNR), Structural Similarity Index Measure (SSIM) and 3D Reconstruction as well. \HLL{Quantitative comparisons} show that equipped with some designs of $\beta$-GANs and the robust $\ell_1$-Autoencoder, one can stabilize the training of GANs and achieve the stat\text{e-}of-th\text{e-}art performance of robust denoising with low SNR data and against possible information contamination. Our proposed methodology thus provides an effective tool for robust denoising of Cryo-EM 2D images, and helpful for 3D structure reconstruction.  

\end{abstract}

\section{\textbf{Introduction}}
\label{intro}
\subsection{\textbf{Robust Denoising in Deep Learning}}
Deep learning technique has rapidly entered into the field of image processing. One of the most popular methods was the Denoising Autoencoder (DA) motivated by~\cite{vincent2008extracting}. It used the reference data to learn a compressed representation (encoder) for the dataset. One extension of DA \HLL{was presented} in~\cite{xie2012image}, which exploited the sparsity regularization and the reconstruction loss to avoid over-fitting. Other developments, such as~\cite{zhang2017beyond}, made use of the residual network architecture to improve the quality of denoised images. In addition,~\cite{agostinelli2013adaptive} combined several sparse Denoising Autoencoder to enhance the robustness under different noise. 

The Generative Adversarial Networks (GAN) recently \HLL{gained} its popularity and provides a promising new approach for image denoising. GAN \HLL{was proposed} by~\cite{goodfellow2014generative}, and was mainly composed of two parts: the generator ($G$: generate the new samples) and the discriminator ($D$: determine whether the samples are real or generated (fake)). Original GAN~(\cite{goodfellow2014generative}) \HLL{aimed to} minimize the Jensen-Shannon (JS) divergence between distributions of the generated samples and the true samples, hence called JS-GAN. Various GANs \HLL{were then studied}, and in particular,~\cite{arjovsky2017wasserstein} proposed the Wasserstein GAN (WGAN), which \HLL{replaced the JS divergence} with Wasserstein distance.~\cite{gulrajani2017improved} further improved the WGAN with the gradient penalty that stabilized the model training. For image denoising problem, GAN could better describe the distribution of original data by exploiting the common information of samples. Consequently, \HLL{GANs were} widely applied in the image denoising problem (\cite{tran2020gan,tripathi2018correction,yang2018low,chen2018image,dong2020optical}).

Recently,~\cite{gao2018robust,gao2019generative} showed that a general family of GANs ($\beta$-GANs, including JS-GAN and TV-GAN) \HLL{enjoyed robust reconstruction} when the data sets contain outliers under Huber contamination models~(\cite{huber1992robust}). In this case, observed samples are drawn from a complex distribution, which is a mixture of contamination distribution and real data distribution. A particular example is provided by Cryo-Electron Microscopy (Cryo-EM) imaging, where the original noisy images are likely contaminated with outliers as broken or non-particles. \HLL{The main challenges of Cryo-EM image denoising are summarized in the subsequent section.}

\subsection{\textbf{Challenges of Cryo-EM Image Denoising}}

The Cryo-Electron Microscopy (Cryo-EM) has become one of the most popular techniques to resolve the atomic structure. In the past, Cryo-EM was limited to large complexes or low-resolution models. Recently the development of new detector hardware has dramatically improved the resolution in Cryo-EM~(\cite{kuhlbrandt2014resolution}), which made Cryo-EM widely used in a variety of research fields. Different from X-ray crystallography\HLL{~(\cite{warren1990x})}, Cryo-EM had the advantage of preventing the recrystallization of inherent water and re‐contamination. Also, Cryo-EM was superior to Nuclear Magnetic Resonance spectroscopy\HLL{~(\cite{wuthrich1986nmr})} in solving macromolecules in the native state. In addition, both X-ray crystallography and Nuclear Magnetic Resonance spectroscopy required large amounts of relatively pure samples, whereas Cryo-EM required much fewer samples~(\cite{bai2015cryo}). For this celebrated development of Cryo-EM for the high-resolution structure determination of biomolecules in solution, the Nobel Prize in Chemistry in 2017 was awarded to three pioneers in this field~(\cite{shen20182017}).

\begin{figure}[htbp]
\centering
\includegraphics[width=6in]{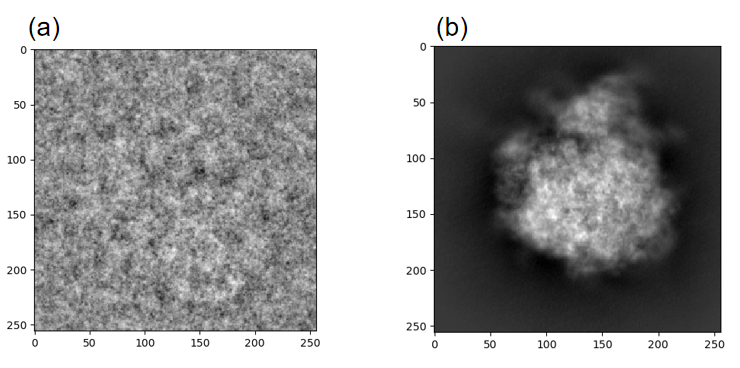}
\caption{\footnotesize (a) a noisy Cryo-EM image (b) a reference image \label{fig:demo}}
\end{figure}

However, it is a computational challenge in processing raw Cryo-EM images, due to heterogeneity in molecular conformations and high noise. Macromolecules in natural conditions are usually heterogeneous, i.e., multiple metastable structures might coexist in the experimental samples (\cite{frank2006three,scheres2016processing}). Such conformational heterogeneity adds extra difficulty to the structural reconstruction as we need to assign each 2D image to not only the correct projection angle but also its corresponding conformation. This imposes a computational challenge that one needs to denoise the Cryo-EM images without losing the key features of their corresponding conformations. Moreover, in the process of generating Cryo-EM images, one needs to provide a view using the electron microscope for samples that are in frozen condition. Thus there are two types of noise: one is from ice, and the other is from the electron microscope. Both of them are significant in contributing high noise in Cryo-EM images and leave a difficulty to the detection of particle structures (Fig. \ref{fig:demo} shows a typical noisy Cryo-EM image with its reference image, which is totally non-identifiable to human eyes). In extreme cases, some experimental images even do not contain any particles, rendering it difficult for particle picking either manually or automatically~(\cite{wang2016deeppicker}). How to achieve robust denoising against such kind of contamination thus becomes a critical problem. Therefore, it is a great challenge to develop robust denoising methods for Cryo-EM images to reconstruct heterogeneous biomolecular structures.

There are a plethora of denoising methods developed in applied mathematics and machine learning that could be applied to Cryo-EM image denoising. Most of them in Cryo-EM are based on unsupervised learning, which don't need any reference image data to learn.~\cite{wang2013zernike} proposed a filtering method based on non-local means, which made use of the rotational symmetry of some biological molecules. Also,~\cite{wei2010optimized} designed the adaptive non-local filter, which made use of a wide range of pixels to estimate the denoised pixel values. Besides,~\cite{xian2018data} compared transform domain filtering method: BM3D (\cite{dabov2007image}) and dictionary learning method: KSVD  (\cite{aharon2006k}) in denoising problem in Cryo-EM. However, all of these \HLL{didn't} work well in low Signal-to-Noise Ratio (SNR) situations. In addition, Covariance Wiener Filtering (CWF) (\cite{bhamre2016denoising}) \HLL{was proposed} for image denoising. \HLL{However,} CWF needed large sample size of data in order to estimate the covariance matrix correctly, although it had an attractive denoising effect. Therefore, a robust denoising method in Cryo-EM images was needed.


\subsection{\textbf{Outline}}
In this chapter, we propose a robust denoising scheme of Cryo-EM images by exploiting joint training of both Autoencoders and a new type of GANs $\beta$-GANs. Our main results are summarized as follows.

\begin{itemize}
    \item Both Autoencoder and GANs help each other for Cryo-EM denoising in low Signal-to-Noise Ratio scenarios. On the one hand, Autoencoder helps stabilize GANs during training, without which the training processes of GANs are often collapsed due to high noise; on the other hand, GANs help Autoencoder in denoising by sharing information in similar samples via distribution learning. For example, WGAN combined with autoencoder often achieve stat\text{e-}of-th\text{e-}art performance due to its ability of exploiting information in similar samples for denoising.
    \item To achieve robustness against partial contamination of samples, one needs to choose both robust reconstruction loss for Autoencoder (e.g., $\ell_1$ loss) and robust $\beta$-GANs (e.g., $(.5,.5)$-GAN or $(1,1)$-GAN\footnote{$\beta$-GAN has two parameters: $\alpha$ and $\beta$, written as $(\alpha, \beta)$-GAN in this chapter.}, which is proved to be robust against Huber contamination), that achieve competitive performance with WGANs in contamination-free scenarios, but do not deteriorate that much with data contamination.  
    \item Numerical experiments are conducted with both a heterogeneous conformational dataset on the Thermus aquaticus RNA Polymerase (RNAP) and a homogenous dataset on the Plasmodium falciparum 80S ribosome dataset (EMPIAR-10028). The experiments on those datasets show the validity of the proposed methodology and suggest that: while WGAN, $(.5,.5)$-GAN, and $(1,1)$-GAN combined with $\ell_1$-Autoencoder are among the best choices in contamination-free cases, the latter two are overall the most recommended for robust denoising.  
\end{itemize}

To achieve the goals above, this chapter is to provide an overview on various developments of GANs with their robustness properties. After that we focus on the application to the challenge of Cryo-EM image robust denoising problem. 

The chapter is structured as follows. In section ``\nameref{sec:background}", we \HLL{provide} a general overview of Autoencoder and GAN. In section ``\nameref{sec:noisy modeling}", we model the tradition denoising problem based on Huber contamination firstly and discuss $\beta$-GAN and its statistics. Finally, we give our algorithm based on combination of $\beta$-GAN and Autoencoder, which is training stable. The evaluation of the algorithm in Cryo-EM date is shown in the section ``\nameref{sec:application}". The section ``\nameref{sec:conclusion}" concludes the chapter. \HLL{In addition, we implement the supplementary experiment in the section ``\nameref{sec:appendix}".}
 
\section{\textbf{Background: Data Representation and Mapping}} \label{sec:background}
Efficient representation learning of data distribution is crucial for many machine learning based model. \HLL{For a set of the real data samples $X$, the classical way to learn the probability distribution of this data ($P_r$) is to find $P_\theta$ by minimizing the distance between $P_r$ and $P_{\theta}$, such as Kullback-Leibler divergence $D_{KL}(P_r||P_{\theta})$. This means we can pass a random variable through a parametric function to generate samples following a certain distribution $P_\theta$ instead of directly estimating the unknown distribution $P_r$.} By varying $\theta$, we can change this distribution and make it close to the real data distribution $P_r$. \HLL{Autoencoder and GANs are two well known methods in data representation.} Autoencoder is good at learning the representation of data with low dimensions with an explicit characterization of $P_\theta$, while GAN offers flexibility in defining the objective function (such as the Jensen-Shannon divergence) by directly generating samples without explicitly formulating $P_\theta$.

\subsection{\textbf{Autoencoder}}
Autoencoder\HLL{~(\cite{baldi2012autoencoders})} is a type of neural network used to learn efficient codings of unlabeled data. It learns a representation (encoding) for a set of data, typically for dimensional reduction by training the network. An Autoencoder has two main parts: encoder and decoder. The encoder maps the input data $x$ ($\in X$) into a latent representation $z$, while the decoder map the latent representation back to the data space:
\begin{align}
    &z\sim \text{Enc}(x)  \\
    &\hat{x} \sim \text{Dec}(z). 
\end{align}
Autoencoders are trained to minimized reconstruction errors, such as: $\mathcal{L}(x,\hat{x})=||x-\hat{x}||^2$.

Various techniques have been developed to improve data representation ability for Autoencoder, such as imposing regularization on the encoding layer:
\begin{align}
    \mathcal{L}(x,\hat{x})+\Omega(h),
\end{align}
where $h$ is the mapping function of the encoding layer, and $\Omega(h)$ is the regularization term. The Autoencoder is good at data denoising and dimension reduction.  

\subsection{\textbf{GAN}}

The Generative Adversarial Network (GAN), firstly proposed by Goodfellow (\cite{goodfellow2014generative}, called JS-GAN), is a class of machine learning framework. The goal of GAN is to learn to generate new data with the same statistics as the training set. Though original GAN is proposed as a form of generative model for unsupervised learning, GAN has proven useful for semi-supervised learning, fully supervised learning and reinforcement learning\HLL{~(\cite{hua2019gan,sarmad2019rl,dai2017good})}.

Although GAN has shown great success in machine learning, the training of GAN is not easy, and is known to be slow and unstable. The problems of GAN\HLL{~(\cite{bau2019seeing,arjovsky2017wasserstein})} include: 
\begin{itemize}
    \item \textit{Hard to achieve Nash equilibrium.} The updating process of the generator and the discriminator models are hard to guarantee a convergence.
    \item \textit{Vanishing gradient.} The gradient update is slow when the discriminator is well trained. 
    \item \textit{Mode collapse.} The generator fails to generate samples with enough representative. 
\end{itemize}

\subsubsection{\textbf{JS-GAN}}
JS-GAN proposed in~\cite{goodfellow2014generative} took Jensen-Shannon (JS) distance to measure the difference between different data distribution. The mathematics expression is follows: 
\begin{align} \label{JSGAN}
    \min\limits_{G}\max\limits_{D}\mathbb{E}_{x\sim P(X),z\sim P(Z)}[\log D(x) + \log (1-D(G(z))],
\end{align}
\HLL{where $G$ is a generator which maps disentangled noise $z\sim P(Z)$ (usually Gaussian $N(0,I)$) to fake image data in an purpose to confuse the discriminator $D$ from real data. The discriminator $D$ is simply a classifier, which makes an effort to distinguish real data from the fake data generated by $G$.} $P(X)$ is the input data distribution. $z$ is noise. $P(Z)$ is the noise distribution, and it is used for data generation. GAN trains the adversarial process by updating generator and the discriminator, where training the generator in succeeding in fooling the discriminator.

\subsubsection{\textbf{WGAN and WGANgp}}
Wasserstein GAN~(\cite{arjovsky2017wasserstein}) replaced the JS distance with the Wasserstein distance:
\begin{equation} \label{eq:wgan}
   \min_{G}\max_{D}\EE_{x\sim P(X),z\sim P(z)}\{D(x) - D(G(z)).
\end{equation}
 In reality, WGAN applied weight clipping of neural network to satisfies Lipschitz condition for discriminator. Moreover,~\cite{gulrajani2017improved} proposed WGANgp based on WGAN, which introduced a penalty in gradient to stabilize the training.
\begin{equation}
   \min_{G}\max_{D}\EE_{(x,z)\sim P(X,z)}\{D(x) - D(G(z))+ \mu \EE_{\tilde{x}} (\Vert \bigtriangledown_{\tilde{x}} D(\tilde{x})\Vert _2 - 1)^2 \}, 
\end{equation}
where $\tilde{x}$ is uniformly sampled along straight lines connecting pairs of the generated and real samples; and
$\mu$ is a weighting parameter. In WGANgp, the last layer of the sigmoid function in the discriminator network is removed. Thus $D$'s output range is the whole real $\RR$, but its gradient is close to $1$ to achieve Lipschitz-$1$ condition. 

\section{\textbf{Robust Denoising Method}} \label{sec:noisy modeling}

\subsection{\textbf{Huber Contamination Noise Model}}
Let $x\in \RR^{d_1\times d_2}$ be a clean image, often called reference image in the sequel. The generative model of noisy image $y\in \RR^{d_1\times d_2}$ under the linear, weak phase approximation~(\cite{bhamre2016denoising}) could be described by 
\begin{equation} \label{eq:forward_model}
    y= a * x + \zeta,  
\end{equation}
where $*$ denotes the convolution operation, $a$
is the point spread function of the microscope convolving with the clean image and $\zeta$ is an additive noise, usually assumed to be Gaussian noise that corrupts the image. In order to remove the noise the microscope brings, traditional Denoising Autoencoder could be exploited to learn from examples $(y_i,x_i)_{i=1,\ldots,n}$ the inverse mapping $a^{-1}$ from the noisy image $y$ to the clean image $x$. 

However, this model is not sufficient in the real case. In the experimental data, the contamination will significantly affect the denoising efficiency if the denoising methods continuously depend on the sample outliers. Therefore we introduce the following Huber contamination model to extend the image formation model \HLL{(see Eq. \eqref{eq:forward_model})}. 

Consider that the pair of reference image and experimental image $(x, y)$ is subject to the following mixture distribution $P_{\epsilon}$: 
\begin{equation} \label{eq:contam_huber}
P_{\epsilon} = (1-\epsilon)P_0 + \epsilon Q, \ \ \ \epsilon\in [0,1],
\end{equation}
a mixture of true distribution $P_0$ of probability $(1-\epsilon)$ and arbitrary contamination distribution $Q$ of probability $\epsilon$. $P_0$ is characterized by Eq. \eqref{eq:forward_model} and $Q$ accounts for the unknown contamination distribution possibly due to ice, broken of data, and so on such that the image sample does not contain any particle information. This is called the Huber contamination model in statistics~\cite{huber1992robust}. Our purpose is that given $n$ samples $(x_i,y_i)\sim P_{\epsilon}$ ($i=1,\ldots,n$), possibly contaminated with unknown $Q$, learn a robust inverse map $a^{-1}(y)$.

\subsection{\textbf{Robust Denoising method}} \label{sec:robust denoise}

Exploiting a neural network to approximate the robust inverse mapping $G_\theta: \RR^{d_1\times d_2} \to \RR^{d_1\times d_2}$. The neural network is parameterized by $\theta\in \Theta$. The goal is to ensure that discrepancy between reference image $x$ and reconstructed image $\widehat{x}=G_\theta(y)$ is small. Such a discrepancy is usually measured by some non-negative loss function: $\ell(x,\widehat{x})$. Therefore, the denoising problem minimizes the following expected loss, 
\begin{equation}
    \arg\min_{\theta\in \Theta}\mathcal{L}(\theta):=\EE_{x,y}[\ell(x,G_{\theta}(y)) ].
\end{equation}

In practice, given a set of training samples $S=\{(x_i,y_i):i=1,\ldots,n\}$, we aim to solve the following empirical loss minimization problem,
\begin{equation}
    \arg\min_{\theta\in \Theta}\widehat{\mathcal{L}}_S(\theta):=\frac{1}{n}\sum_{i=1}^n \ell(x_i,G_{\theta}(y_i)).
\end{equation}
the following choices of loss functions are considered:
\begin{itemize}
    \item ({\textbf{$\ell_2$-Autoencoder}}) $\ell(x,\widehat{x})=\frac{1}{2}\|x-\widehat{x}\|_2^2:=\frac{1}{2}\sum_{i,j} (x_{ij}-\widehat{x}_{ij})^2$, or $\EE\ell(x,\widehat{x})=D_{KL}(p(x)\|q(\widehat{x}_\theta))$ equivalently, where $\widehat{x}_\theta\sim \mathcal{N}(x,\sigma^2 I_D)$;
    
    \item ({\textbf{$\ell_1$-Autoencoder}}) $\ell(x,\widehat{x})=\|x-\widehat{x}\|_1:=\sum_{i,j} |x_{ij}-\widehat{x}_{ij}|$, or $\EE\ell(x,\widehat{x})=D_{KL}(p(x)\|q(\widehat{x}_\theta)) $ equivalently, where $\widehat{x}_\theta\sim \Lap(x,b)$;
    \item ({\textbf{Wasserstein-GAN}})
    $\ell(x,\widehat{x})=W_1(p(x),q_\theta(\widehat{x}))$, where $W_1$ is the 1-Wasserstein distance between distributions of $x$ and $\widehat{x}$;
    
    \item ({\textbf{$\beta$-GAN}})
    $\ell(x,\widehat{x})=D(p(x)\|q_\theta(\widehat{x}))$, where $D$ is some divergence function to be discussed below between distributions of $x$ and $\widehat{x}$.
\end{itemize}

Both the $\ell_2$ and $\ell_1$ losses consider the reconstruction error of $G_\theta$. The $\ell_2$-loss above is equivalent to assume that $G_\theta(y|x)$ follows a Gaussian distribution $\mathcal{N}(x,\sigma^2 I_{D})$, and the $\ell_1$-loss instead assumes a Laplacian distribution centered at $x$. As a result, the $\ell_2$-loss pushes the reconstructed image $\widehat{x}$ toward mean by averaging out the details and thus blurs the image. On the other hand, the $\ell_1$-loss pushes $\widehat{x}$ toward the coordinat\text{e-}wise median, keeping the majority of details while ignoring some large deviations. It improves the reconstructed image, and more robust than the $\ell_2$ loss against large outliers. Although $\ell_1$-Autoencoder has a more robust loss than $\ell_2$, both of them are not sufficient to handle the contamination. In the framework of the Huber contamination model (\HLL{Eq. \eqref{eq:contam_huber}}), $\beta$-GAN is introduced below.

\subsection{\textbf{Robust Recovery via $\beta$-GAN}}

Recently~\cite{gao2018robust,gao2019generative} came up with a more general form of $\beta$-GAN. It aims to solve the following minimax optimization problem to find the $G_\theta$, 
\begin{equation}\label{minmax equation}
    \min_{G_\theta}\max_{D}\EE[S(D(x),1) +S(D(G_\theta(y)), 0)], 
\end{equation}
 where $S(t,1) = -\int_t^1 c^{\alpha-1}(1-c)^{\beta} dc$, $S(t,0) = -\int_0^t c^{\alpha}(1-c)^{\beta-1} dc$, $\alpha, \beta\in [-1,1]$. For simplicity, we denote this family with parameters $\alpha, \beta$ by ($\alpha,\beta$)-GAN in this chapter. 
 
The family of ($\alpha,\beta$)-GAN includes many popular members. For example, when $\alpha = 0, \beta = 0$, it becomes the JS-GAN~(\cite{goodfellow2014generative}), which aims to solve the minmax problem~ \HLL{(Eq. \eqref{JSGAN})} whose loss is the Jensen-Shannon divergence.
When $\alpha = 1, \beta =1$ the loss is a simple mean square loss; when $\alpha =-0.5, \beta = -0.5$, the loss is boost score.

However, the Wasserstein GAN (WGAN) is not a member of this family. By formally taking $S(t,1)=t$ and $S(t,0)=-t$, we could derive the the type of WGAN as \HLL{Eq.~\eqref{eq:wgan}.}

\subsubsection{\textbf{Robust Recovery Theory}} \label{robust theo}
Extending the traditional image generative model to a Huber contamination model and exploit the $\beta$-GAN toward robust denoising under unknown contamination. Below includes a brief introduction to robust $\beta$-GAN, which achieves provable robust estimate or recovery under Huber contamination model. Recently, Gao establishes the statistical optimality of $\beta$-GANs for robust estimate of mean (location) and convariance (scatter) of the general elliptical distributions (\cite{gao2018robust},~\cite{gao2019generative}). Here we introduce the main results.

\begin{definition}[Elliptical Distribution]
A random vector $X \in \RR^p$ follows an elliptical distribution if and only if it has the representation $X = \theta + \xi AU$, where $\theta \in \RR^p$ and $A \in \RR^{p \times r}$ are model parameters.
The random variable U is distributed uniformly on the unit sphere \{$u \in \RR^r: \parallel u \parallel = 1$\} and $\xi \geq 0$ is a random variable in $\RR$ independent of $U$. The vector $\theta$ and the matrix $\Sigma = AA^T$ are called the location and the scatter of the elliptical distribution.
\end{definition}
Normal distribution is just a member in this family characterized by mean $\theta$ and covariance matrix $\Sigma$. Cauchy distribution is another member in this family whose moments do not exist. 

\begin{definition}[Huber contamination model]
 $X_1,..., X_n \overset{\text{iid}}{\sim} (1-\epsilon)P_{ell} + \epsilon Q$, where we consider the $P_{ell}$ an elliptical distribution in its canonical form.
\end{definition}

A more general data generating process than Huber contamination model is called the strong
contamination model below, as the $TV$-neighborhood of a given elliptical distribution $P_{ell}$:
\begin{definition}[Strong contamination model] \label{def:TV-ambiguity}
$X_1,..., X_n \overset{\text{iid}}{\sim} P$, for some $P$ satisfying 
$$TV(P, P_{ell}) < \epsilon .$$ 
\end{definition}
\begin{definition}[Discriminator Class]
Let $\sig(x)=\frac{1}{1+e^{-x}}$, $\ramp(x)=\max(\min(x+1/2,1),0)$, and $\relu(x)=\max(x,0)$. Define a general discriminator class of deep neural nets: firstly define the 
a ramp bottom layer
\begin{equation}
    \G_{\ramp} = {g(x) = \ramp(u^tx+b), u\in \RR^p, b \in \RR}.
\end{equation} 
\\Then, with $\G_1(B) = \G_{\ramp}$, inductively define
\begin{equation}
\G_{l+1}(B) = \Bigg \{ g(x)=\relu \bigg (\sum_{h\geq1} v_h g_h(x) \bigg): \sum_{h\geq1}|v_h| \leq B, g_h \in \G_l(B) \Bigg \}.
\end{equation}
Noted that the neighboring two layers are connected via ReLU activation functions. Finally,
the network structure is defined by:
\begin{equation} \label{discri class}
   \D^L(\kappa, B) = \Bigg \{ D(x) = \sig \bigg( \sum_{j\geq1}w_jg_j(x) \bigg): \sum_{j\geq1}|w_j|\leq \kappa, g_j \in \G_L(B) \Bigg \}.
\end{equation}
\HLL{This is a network architecture consisting of} $L$ hidden layers.
\end{definition}

Now consider the following $\beta$-GAN induced by a proper scoring rule $S:[0,1]\times \{0,1\} \to \RR$ with the discriminator class above: 
\begin{equation} \label{estimation para}
   (\hat{\theta}, \hat{\Sigma}) = \arg\min_{(\theta, \Sigma)}\max_{D \in \D^L(\kappa,B)}\frac{1}{n}\sum_{i=1}^nS(D(x_i),1) +\EE_{x \sim P_{ell}(\Theta, \Sigma)}S(D((x)), 0).
\end{equation}
The following theorem shows that such a $\beta$-GAN may give a statistically optimal estimate of location and scatter of the general family of Elliptical distributions under strong contamination models. 
\begin{thm}[Gao-Yao-Zhu~(\cite{gao2019generative})]\label{thm1}
Consider the $(\alpha,\beta)$-GANs with $|\alpha-\beta|<1$. The discriminator class $D=\D^L(k, B)$ is specified by Eq. \eqref{discri class}. Assume $\frac{p}{n} + \epsilon ^2 \leq c$ for some sufficiently small constant $c > 0$. Set $1 \leq L = O(1), 1 \leq B = O(1)$, and $\kappa = O(\sqrt{\frac{p}{n}} + \epsilon)$. Then for any $X_1,...X_n \overset{\text{iid}}{\sim} P$, for some $P$ satisfying $TV(P, P_{ell}) < \epsilon$ with small enough $\epsilon$, we have:
\begin{equation} 
\begin{split}
&\| \hat{\theta} -\theta \|^2 < C (\frac{p}{n} \vee \epsilon ^2), \\
&\| \hat{\Sigma} -\Sigma \|_{op}^2 < C (\frac{p}{n} \vee \epsilon ^2),
\end{split}
\end{equation}
with probability at least $1 - e^{C'(p+n\epsilon^2)}$ (universal constants $C$ and $C'$) uniformly over all $\theta \in R^p$ and all $\|\Sigma\|_{op}\leq M$. 
\end{thm} 

The theorem established that for all $|\alpha-\beta|<1$, $(\alpha,\beta)$-GAN family is robust in the sense that one can learn a distribution $P_{ell}$ from contaminated distributions $P_{\epsilon}$ such that $TV(P_{\epsilon},P_{ell})<\epsilon$, which includes Huber contamination model as a special case. Therefore a $(\alpha,\beta)$-GAN with suitable choice of network architecture, can robustly learn the generative model from arbitrary contamination $Q$ when $\epsilon $ is small (e.g. no more than $1/3$). 

\HLL{In the current case, the denoising autoencoder network is modified to $G_\theta(y)$,  providing us an universal approximation of the location (mean) of the inverse generative model as Eq. \eqref{eq:forward_model},} where the noise can be any member of the elliptical distribution. Moreover, the discriminator is adapted to the image classification problem in the current case. Equipped with this design, the proposed $(\alpha,\beta)$-GAN may help enhance the denoising Autoencoder robustness against unknown contamination, e.g. the Huber contamination model for real contamination in the image data. The experimental results in fact confirms the efficacy of such a design.

In addition, Wasserstein GAN (WGAN) is not a member of this $\beta$-GAN family. Compared to JS-GAN, WGAN aims to minimize the Wasserstein distance between the sample distribution and the generator distribution. Therefore, WGAN is not robust in the sense of contamination models above as arbitrary $\epsilon$ portion of outliers can be far away from the main distribution $P_0$ such that the Wasserstein distance is arbitrarily large.



\subsection{\textbf{Stablized Robust Denoising by Joint Autoencoder and $\beta$-GAN}}\label{sec:joint algorithm}
Although $\beta$-GAN can robustly recover model parameters with contaminated samples, as a zero-sum game involving a non-convex-concave minimax optimization problem, training GANs is notoriously unstable with typical cyclic dynamics and possible mode collapse entrapped by local optima (\cite{arjovsky2017wasserstein}). However, in this section we show that the introduction of Autoencoder loss is able to stabilize the training and avoid the mode collapse. In particular, Autoencoder can help stabilize GAN during training, without which the training processes of GAN are often oscillating and sometimes collapsed due to the presence of high noise.

\HLL{Compared with the autoencoder, $\beta$-GAN can further help denoising by exploiting common information in similar samples during distribution training.} In GAN, the divergence or Wasserstein distance between the reference image set and the denoised image set is minimized. The similar images can therefore help boost signals for each other. 

For these considerations, a combined loss is proposed with both $\beta$-GAN and Autoencoder reconstruction loss,
\begin{equation}
    \widehat{\mathcal{L}}_{GAN}(x,\widehat{x})+ \lambda \|x-\widehat{x}\|_p^p,
\end{equation}
where $p\in\{1,2\}$ and $\lambda\geq 0$ is a trad\text{e-}off parameter for $\ell_p$ reconstruction loss. Algorithm \ref{mainalg} summarizes the procedure of joint training of Autoencoder and GAN, which will be denoted as ``GAN$+\ell_p$" in the experimental section depending on the proper choice of GAN and $p$. The main algorithm is shown in Algorithm \ref{mainalg}.

\begin{algorithm} \footnotesize
\caption{Joint training of $(\alpha,\beta)$-GAN and $\ell_p$-Autoencoder.}\label{mainalg}
{\bf Input:} \\
1. $(\alpha,\beta)$ for $S(t,1) = -\int_t^1 c^{\alpha-1}(1-c)^{\beta}$dc, $S(t,0) = -\int_0^t c^{\alpha} (1-c)^{\beta-1}$dc  \\ \hspace*{0.05in}  or $S(t,1)=t$, $S(t,0)=-t$ for WGAN \\
2. $\lambda$ regularization parameter of the $\ell_p$-Autoencoder \\
3. $k_d$ number of iterations for discriminator, $k_g$ number of iterations for generator\\
4. $\eta_d$ learning rate of discriminator, $\eta_g$ learning rate of generator\\
5. $\omega$ weights of discriminator, $\theta$ weights of generator

\begin{algorithmic}[1]
\FOR{number of training iterations}{}
\STATE $\bullet$ Sample minibatch of $m$ examples $\{(x^{(1)},y^{(1)}),\ldots,(x^{(m)},y^{(m)})\}$ from referenc\text{e-}noisy image pairs.
\FOR{$k =1,2...,k_d$}{}
\STATE $\bullet$ Update the discriminator by gradient ascent:
\STATE $g_{\omega} \xleftarrow{}  \frac{1}{m} \sum_{i=1}^m \nabla_\omega [ S(D_\omega(x_i),1) + S(D_\omega(G_\theta(y_i)),0)+ \mu  (\Vert \bigtriangledown_{\tilde{x}} D_\omega(\tilde{x}_i)\Vert _2 - 1)^2 ]$ \\ \ \ \ \ \ \ where $\mu>0$ for WGANgp only;
\STATE $\omega \xleftarrow{} \omega + \eta_d g_{\omega} $
\ENDFOR
\FOR{$k =1,2...,k_g$}{}
\STATE $\bullet$ Update the generator by gradient descent:
\STATE $g_{\theta} \xleftarrow{} \frac{1}{m} \sum_{i=1}^m \nabla_\theta [ S(D_\omega(G_\theta(y_i)),0) + \lambda \lvert G_\theta(y_i) - x_i \rvert^p]$, $p\in \{1,2\}$  ;
\STATE $\theta \xleftarrow{} \theta - \eta_g g_{\omega} $
\ENDFOR 
\ENDFOR
\end{algorithmic}
\textbf{Return}:Denoised image: $\widehat{x}_i=G_\theta(y_i)$
\label{algorithm}
\end{algorithm}

\subsubsection{\textbf{Stability of combining Autoencoder into GAN}} \label{sec:stability}
We illustrate that Autoencoder is indispensable to GANs in stabilizing the training in the joint training of Autoencoder and GAN scheme. 

\begin{figure}[htbp]
\centering
\includegraphics[width=6in, height=65mm]{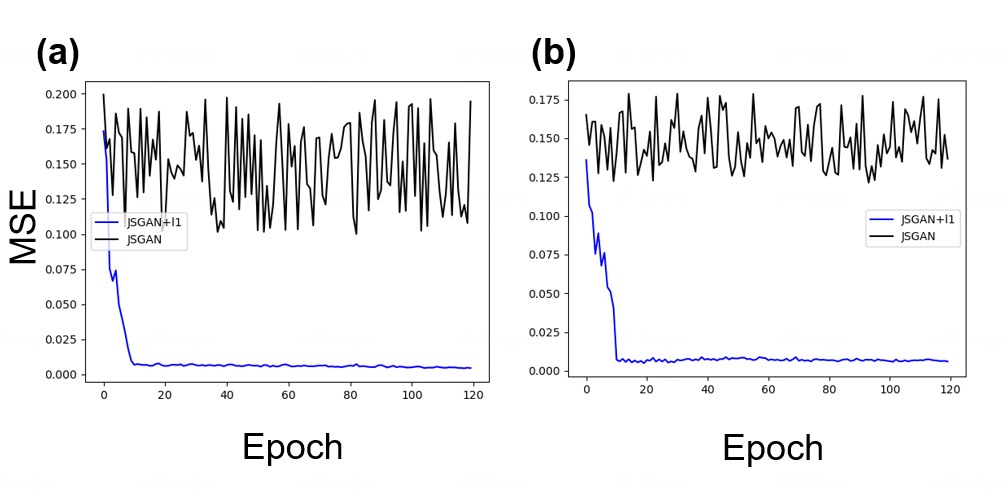}
\caption{Comparison between JS-GAN (black) and joint JS-GAN-$\ell_1$ Autoencoder (blue). \HLL{(a) and (b) are the change of MSE in training and testing data.} Joint training of JS-GAN-$\ell_1$ Autoencoder is much more stable than pure JS-GAN training that oscillates a lot. }
\label{fig: l1-nonl1}
\end{figure}

As an illustration, Fig. \ref{fig: l1-nonl1} shows the comparison of training a JS-GAN and a joint JS-GAN + $\ell_1$-Autoencoder. Training and test mean square error curves are plotted against iteration numbers in the RNAP data under $SNR = 0.1$ as Fig. \ref{fig: l1-nonl1}. It shows that JS-GAN training suffers from drastic oscillations while joint training of JS-GAN + $\ell_1$-Autoencoder exhibits a stable process. In fact, with the aid of Autoencoder here, one does not need the popular ``$\log D$ trick" in JS-GAN.

\section{\textbf{Application: Robust Denoising of Cryo-EM Images}} \label{sec:application}

\subsection{\textbf{Datasets}}
\subsubsection{\textbf{RNAP: Simulation Dataset}}
\begin{figure}[htbp]
\centering
\includegraphics[width=6in]{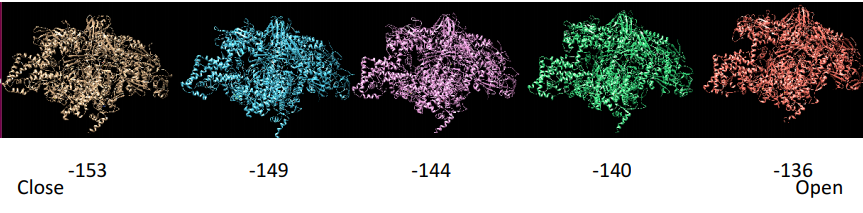}
\caption{Five conformations in RNAP heterogeneous dataset, from left to right are close conformation to open conformation of different angles. \label{fig:5conf}}
\end{figure}

We design a conformational heterogeneous dataset obtained by simulations. We use \textit{Thermus aquaticus} RNA Polymerase (RNAP) in complex with $\sigma^A$ factor (\textit{Taq} holoenzyme) for our dataset. RNAP is the enzyme that transcribes RNA from DNA (transcription) in the cell. During the initiation of transcription, the holoenzyme must bind to the DNA, then separate the doubl\text{e-}stranded DNA into singl\text{e-}stranded (\cite{browning2004regulation}). \textit{Taq} holoenzyme has a crab-claw like structure, with two flexible domains, the clamp and $\beta$ pincers. The clamp, especially, has been suggested to play an important role in the initiation, as it has been captured in various conformations by CryoEM during initiation (\cite{chen2020stepwise}). Thus, we focus on the movement of the clamp in this study. To generate the heterogeneous dataset, we start with two crystal structures of \textit{Taq} holoenzyme, which vary in their clamp conformation, open  (PDB ID: 1L9U (\cite{murakami2002structural})) and closed (PDB ID: 4XLN (\cite{bae2015structure})) clamp. For the closed clamp structure, we remove the DNA and RNA in the crystal structure, leaving only the RNAP and $\sigma^A$ for our dataset. The \textit{Taq} holoenzyme has about 370 kDa molecular weight. We then generate the clamp intermediate structures between the open and closed clamp using multipl\text{e-}basin coars\text{e-}grained (CG) molecular dynamic (MD) simulations (\cite{okazaki2006multiple,kenzaki2011cafemol}). CG-MD simulations simplify the system such that the atoms in each amino acid are represented by one particle. The structures from CG-MD simulations are refined back to all-atom or atomic structures using PD2 ca2main (\cite{moore2013high}) and SCRWL4 (\cite{krivov2009improved}). Five structures with equally-spaced clamp opening angle are chosen for our heterogeneous dataset (shown in Fig. \ref{fig:5conf}).  Then, we convert the atomic structures to $128 \times 128 \times 128$ volumes using \texttt{Xmipp} package (\cite{marabini1996xmipp}) and generate the 2D projections with an image size of $128 \times 128$ pixels. We further contaminate those clean images with additive Gaussian noise at different Signal-to-Noise Ratio (SNR): $SNR =0.05$. The SNR is defined as the ratio of signal power and the noise power in the real space. For simplicity, we did not apply the contrast transfer function (CTF) to the datasets, and all the images are centered. Fig. \ref{fig:5conf} shows the five \HLL{conformation pictures.}

Training data size is $25000$ paired images(noisy and reference images), Test data to calculate the MSE, PSNR and SSIM is another $1500$ paired images.

\subsubsection{\textbf{EMPIAR-10028: Real Dataset}}
\begin{figure}[htbp]

\includegraphics[width=6in]{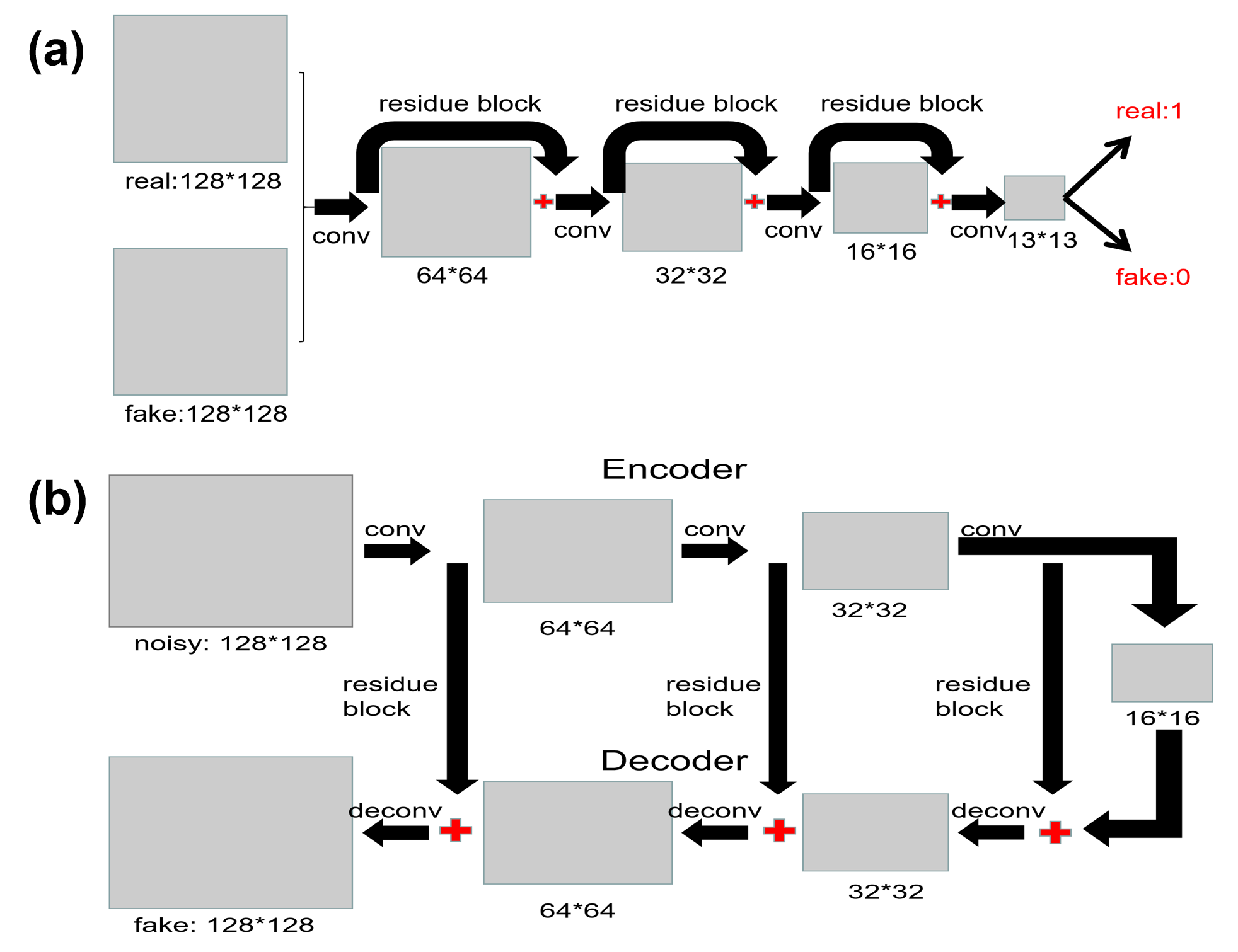}

\caption{The architectures of (a) discriminator $D$ and (b) generator $G$, which borrow the residue structure. The input image size ($128 \times 128 $) here is adapted to RNAP dataset, while input image size of EMPIAR-10028 dataset is $256 \times 256$.  \label{fig:architecture}}
\end{figure}

 This is a real-world experimental dataset that was firstly studied in: the Plasmodium falciparum 80S ribosome dataset (EMPIAR-10028) (\cite{wong2014cryo}). They recover the Cryo-EM structure of the cytoplasmic ribosome from the human malaria parasite, \textit{Plasmodium falciparum}, in complex with emetine, an anti-protozoan drug, at $3.2 \mathring{ A}$ resolution. Ribosome is the essential enzyme that translates RNA to protein molecules, the second step of central dogma. The inhibition of ribosome activity of \textit{Plasmodium falciparum} would effectively kill the parasite (\cite{wong2014cryo}). We can regard this dataset to have homogeneous property. This dataset contains $105247$ noisy particles with an image size of $360 \times 360$ pixels. In order to decrease the complexity of the computing, we pick up the center square of each image with a size of $256 \times 256$, since the surrounding area of the image is entirely useless that does not lose information in such a preprocessing. Then the $256 \times 256$ images are fed as the input of the $G_\theta$-network (Fig. \ref{fig:architecture}). Since the GAN-based method needs clean images as reference, we prepare their clean counterparts in the following way: we first use cryoSPARC1.0 (\cite{punjani2017cryosparc}) to build a $3.2 A$ resolution volume and then rotate the 3D-volume by the Euler angles obtained by cryoSPARC to get projected 2D-images. The training data size we pick is $19500$, and the test data size is $500$.

\subsection{\textbf{Evaluation Method}}
\label{sec: evaluation}
We exploit the following three metrics to determine whether the denoising result is good or not. They are the Mean Square Error (MSE), the Peak Signal-to-Noise Ratio (PSNR) and the Structural Similarity Index Measure (SSIM). 
\begin{itemize}
    \item (\textbf{MSE}) 
    For images of size $d_1\times d_2$, the Mean Square Error (MSE) between the reference image $x$ and the denoised image $\widehat{x}$ is defined as, 
    \begin{align*}
    \mbox{MSE} := \frac{1}{d_1d_2}\sum_{i=1}^{d_1}\sum_{j=1}^{d_2} (x(i,j)-\widehat{x}(i,j))^2.
    \end{align*}
    The smaller is the MSE, the better the denoising result is.
    
    \item (\textbf{PSNR)} 
    Similarly, the Peak Signal-to-Noise Ratio (PSNR) between the reference image $x$ and the denoised image $\widehat{x}$ whose pixel value range is $[0,t]$ (1 by default), is defined by 
    \begin{align*}
    \mbox{PSNR} :=10\log_{10}\frac{t^2}{\frac{1}{d_1d_2}\sum_{i=1}^{d_1}\sum_{j=1}^{d_2} (x(i,j)-\widehat{x}(i,j))^2}.
    \end{align*}
    The larger is the PSNR, the better the denoising result is. 
    \item (\textbf{SSIM)} 
    The third criterion is the Structural Similarity Index Measure (SSIM) between reference image $x$ and denoised image $\widehat{x}$ is defined in~\cite{wang2004image}, 
    \begin{align*}
    \mbox{SSIM}= \frac{(2\mu_x \mu_{\widehat{x}} +c_1)(2 \sigma_x \sigma_{\widehat{x}} +c_2)(\sigma_{x \widehat{x}} +c_3)}{(\mu_x ^2 + \mu_{\widehat{x}}^2 + c_1)(\sigma_x^2 + \sigma_{\widehat{x}}^2 + c_2)(\sigma_{x} \sigma_{\widehat{x}} +c_3)}.
    \end{align*}
    where $\mu_x$ ($\mu_{\widehat{x}}$) and $\sigma_x$ ($\sigma_{\widehat{x}}$) are the mean and variance of $x$ ($\widehat{x}$), respectively, $\sigma_{x \widehat{x}}$ is covariance of $x$ and $\widehat{x}$, $c_1= {K_1L}^2$, $c_2={K_2L}^2$, $c_3=\frac{c_2}{2} $ three variables to stabilize the division with weak denominator ($K_1 =0.01$, $K_2=0.03$ by default), $L$ is  the dynamic range of the pixel-value (1 by default). The value SSIM of lies in $[0,1]$, where the closer it is to 1, the better the result is. 
    
\end{itemize}

Although these metrics are widely used in image denoising, they might not be the best metrics for Cryo-EM images. In Appendix ``\nameref{sec:lambda}", it shows an example that the best-reconstructed images perhaps do not meet the best MSE/PSNR/SSIM metrics.

In addition to these metrics, we consider the 3D reconstruction based on denoised images. Particularly, we take the 3D reconstruction by RELION to validate the denoised result. The procedure of our RELION reconstruction is follows: firstly creating the 3D initial model, then doing 3D classification, followed by operating 3D auto-refine. Moreover, for heterogeneous conformations in simulation data, we further turn the denoising results into a clustering problem to measure the efficacy of denoising methods, whose details will be discussed in Appendix ``\nameref{sec:cluster}''.

\subsection{\textbf{Network Architecture and Hyperparameter}} \label{sec:net}

In the experiments of this chapter, the best results come from the ResNet architecture (\cite{su2018generative}) shown in Fig. \ref{fig:architecture}, which has been successfully applied to study biological problems such as predicting protein-RNA binding. The generator in such GANs exploits the Autoencoder network architecture, while the discriminator is a binary classification ResNet. In Appendix ``\nameref{sec:convnet}" and ``\nameref{sec:PGGAN}", we also discuss a Convolutional Network without residual blocks and the PGGAN (\cite{karras2018progressive}) strategy with their experimental results, respectively. 

 We chose Adam  (\cite{kingma2015adam}) for the Optimization. The learning rate of the discriminator is $\eta_d=0.001$, and the learning rate of the generator is $\eta_g=0.01$. We choose $m=20$ as our batch size,  $k_d=1$, and $k_g=2$ in Algorithm \ref{algorithm}.

For $(\alpha,\beta)$-GAN, we reports two types of choices: (1) $\alpha=1$, $\beta= 1$; (2) $\alpha=0.5, \beta=0.5$ since they show the best results in our experiments, while the others are collected in Appendix ``\nameref{sec:betagans}". For WGAN, the gradient penalty with parameter $\mu=10$ is used to accelerate the speed of convergence and hence the algorithm is denoted as WGANgp below. The trad\text{e-}off (regularization) parameter of $\ell_1$ or $\ell_2$ reconstruction loss is set to be $\lambda=10$ through out this section, while an ablation study on varying $\lambda$ is discussed in Appendix ``\nameref{sec:lambda}". 


\subsection{\textbf{Results for RNAP}} \label{sec:RNAP}
\subsubsection{\textbf{Denoising without contamination}}
\begin{figure}[htbp]
\centering
\includegraphics[width=6in]{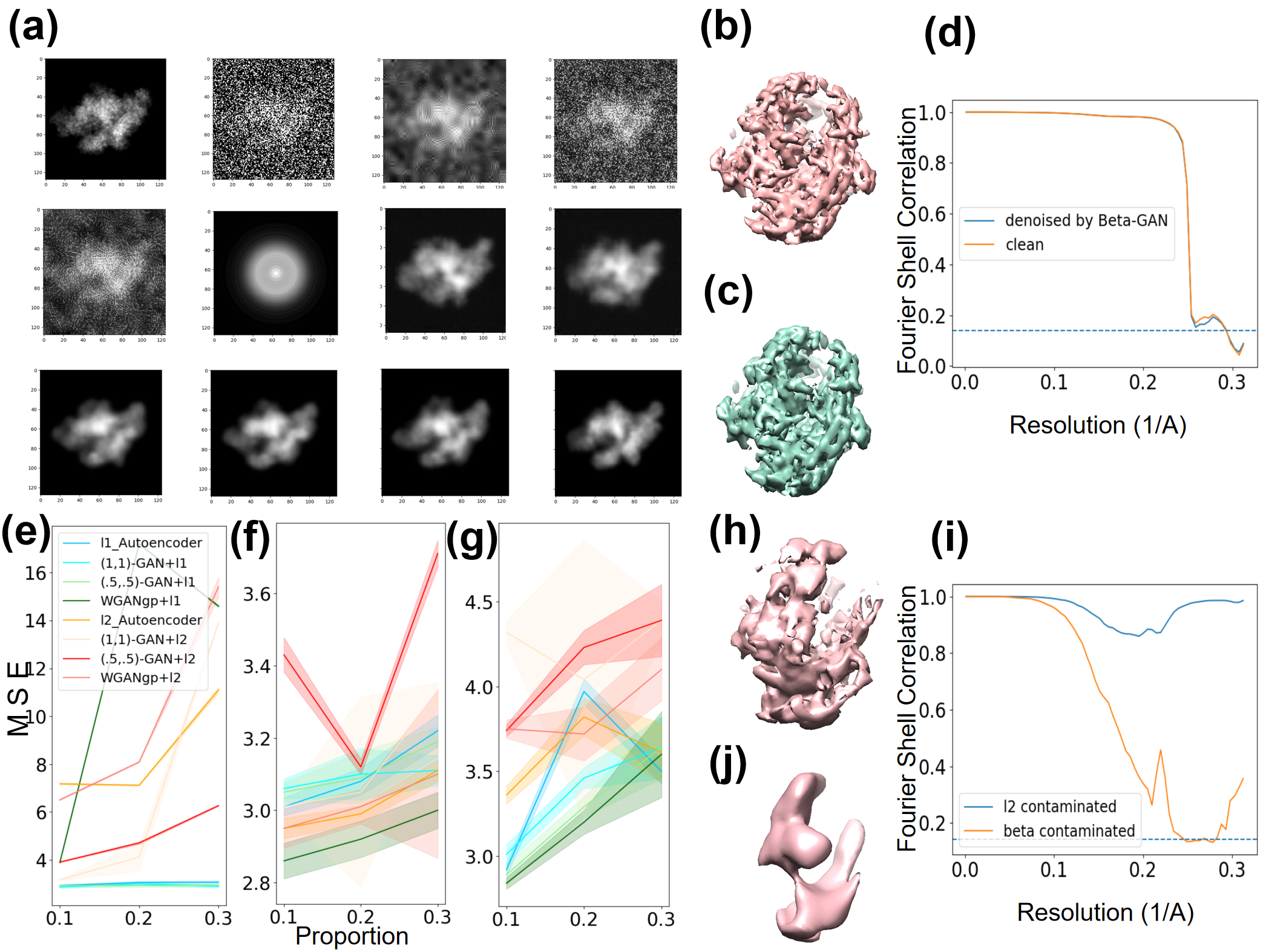}
\caption{Results for RNAP dataset. (a) is denoised images in different denoised methods (from left to right, top to bottom): Clean, Noisy, BM3D, KSVD, Non-local means, CWF, $\ell_1$-Autoencoder, $\ell_2$-Autoencoder, (1,1)-GAN + $\ell_1$, $(0,0)$-GAN + $\ell_1$, $(.5,.5)$-GAN + $\ell_1$ and WGANgp + $\ell_1$. (b) and (c) are reconstruction of clean images and $(.5,.5)$-GAN + $\ell_1$ denoised images. (d) is FSC curve of (b) and (c). (e), (f) and (g) are 
robustness tests of various methods under $\epsilon\in \{0.1, 0.2, 0.3\}$-proportion contamination in three types of contamination: (e) Type A: replacing the reference images with random noise; (f) Type B: replacing the noisy images with random noise; (g) Type C: replacing both with random noise. (h) and (j) are reconstructions of images with $(.5,.5)$-GAN + $\ell_1$ and $\ell_2$-Autoencoder under type A contamination, respectively, where $\ell_2$-Autoencoder totally fails but $(.5,.5)$-GAN + $\ell_1$ is robust. (i) shows FSC curves of (h) and (j).}
\label{fig4}
\end{figure}

In this part, we attempt to denoise the noisy image without the contamination (i.e., $\epsilon=0$ in \HLL{Eq. \eqref{eq:contam_huber}}). In order to present the advantage of GAN, we compare the denoising result in different methods. Table \ref{tbl: sim_denoise} shows the MSE and PSNR of different methods in SNR $0.05$ and $0.1$. We recognize the traditional methods such as KSVD, BM3D, Non-local mean, and CWF can remove the noise partially and extract the general outline, but they still leave the unclear piece. However, deep learning methods can perform much better. Specifically, we observe that GAN-based methods, especially WGANgp $+\ell_1$ loss and $(.5,.5)$-GAN $+\ell_1$ loss, perform better than denoising Autoencoder methods, which only optimizes $\ell_1$ or $\ell_2$ loss. The adversarial process inspires the generation process, and the additional $\ell_1$ loss optimization speeds up the process of generation towards reference images. Notably, WGANgp and $(5,.5)$- or $(1,1)$-GANs are among the best methods, where the best mean performance up to one standard deviation are all marked in bold font. Specifically, compared with $(.5,.5)$-GAN, the WGANgp get better PSNR and SSIM in SNR 0.1; the $(.5,.5)$-GAN shows the advantage in PSNR and SSIM in SNR $0.05$ while $(1,1)$-GAN is competitive within one standard deviation. Also, Fig. \ref{fig4}(a) presents the denoised images of denoising methods in SNR $0.05$. For the convenience of comparison, we choose a clear open-conformation \HLL{(the rightmost conformation of Fig. \ref{fig:5conf})} to present, and the performances show that WGANgp and $(\alpha,\beta)$-GAN can grasp the ``open" shape completely and derive the more explicit pictures than other methods.

What's more, in order to test the denoised results of $\beta$-GAN, we reconstruct the 3D volume by RELION in 200000 images of SNR 0.1, which are denoised by $(.5,.5)$-GAN + $\ell_1$. The value of pixel size, amplitude contrast, spherical aberration and voltage are 1.6, 2.26, 0.1 and 300. For the other terms, retaining the default settings in RELION software. Fig. \ref{fig4}(b) and (c) separately show the 3D volume recoverd by clean images and denoised images. Also, the related FSC curves are shown in Fig. \ref{fig4}(d). Specifically, the blue curve, which represents the denoised images in $(.5,.5)$-GAN + $\ell_1$ is closed to red curves representing the clean images. We use the 0.143 cutoff criterion in literature (the resolution as Fourier shell correlation reaches 0.143, shown by dash lines in Fig. \ref{fig4}(d)) to choose the final resolution: 3.39$\mathring{A}$. The structure recovered by $(.5,.5)$-GAN + $\ell_1$ and FSC curve are as good as the original structure, which illustrates that the denoised result of $\beta$-GAN can identify the details of image and be helpful in 3D reconstruction.

\begin{table}[t]\tiny
\renewcommand\arraystretch{1.5}
\centering
\caption{Denoising result without contamination in simulated RNAP dataset: MSE, PSNR and SSIM of different models, such as BM3D (\cite{dabov2007image}), KSVD (\cite{aharon2006k}), Non-local means (\cite{wei2010optimized}), CWF (\cite{bhamre2016denoising}), DA and GAN-based methods. }
\begin{tabular}{|c|c|c|c|c|c|c|}
\hline
 & \multicolumn{2}{c|}{MSE} & \multicolumn{2}{c|}{PSNR}& \multicolumn{2}{c|}{SSIM}  \\ \hline
  Method/SNR &  0.1 &0.05  &0.1 & 0.05 &0.1 & 0.05 \\   
\hline
BM3D  & 3.52\text{e-}2 (7.81\text{e-}3)& 5.87\text{e-}2(9.91\text{e-}3) & 14.54(0.15) & 12.13(0.14)&0.20(0.01)&0.08(0.01)\\ \hline
KSVD  & 1.84\text{e-}2(6.58\text{e-}3) & 3.49\text{e-}2(7.62\text{e-}3)& 17.57(0.16)& 14.61(0.14) &0.33(0.01)&0.19(0.01)  \\ \hline
Non-local means & 5.02\text{e-}2(5.51\text{e-}3)&5.81\text{e-}2(8.94\text{e-}3)&13.04(0.50)&12.40(0.65)&0.18(0.01)&0.09(0.01)  \\ \hline
CWF  & 2.53\text{e-}2(2.03\text{e-}3) & 9.28\text{e-}3(8.81\text{e-}4)& 16.06(0.33)& 20.31(0.41)&0.25(0.01)&0.08(0.01)  \\ \hline
$\ell_2$-Autoencoder\tablefootnote{$\ell_2$-Autoencoder represents $\ell_2$ loss}  & 3.13\text{e-}3(7.97\text{e-}5) & 4.02\text{e-}3(1.48\text{e-}4)& 25.10(0.11)& 23.67(0.77)&0.79(0.02)&\textbf{0.79(0.01)} \\ \hline
$\ell_1$-Autoencoder\tablefootnote{$\ell_1$-Autoencoder  represents $\ell_1$ loss}  & 3.16\text{e-}3(7.05\text{e-}5) & 4.23\text{e-}3(1.32\text{e-}4)& 25.05(0.09)& 23.80(0.13) &0.77(0.02)&0.76(0.01)\\ \hline
$(0,0)$-GAN + $\ell_1$ \tablefootnote{GAN + $\ell_1$  represents adding $\ell_1$ regularization in GAN generator loss}  & 3.06\text{e-}3(5.76\text{e-}5) &\textbf{ 4.02\text{e-}3(5.67\text{e-}4)} & 25.25(0.04) & 24.00(0.06)&0.78(0.03)&\textbf{0.78(0.03)}     \\ \hline
WGANgp + $\ell_1$  & \textbf{2.95\text{e-}3(1.41\text{e-}5)} & \textbf{4.00\text{e-}3(8.12\text{e-}5)} & \textbf{25.42(0.04)} & \textbf{24.06(0.05)}&\textbf{0.83(0.02)}&\textbf{0.80(0.03) }\\ \hline
$(1,1)$-GAN + $\ell_1$ & 2.99\text{e-}3(3.51\text{e-}5) & \textbf{4.01\text{e-}3(1.54\text{e-}4)} &  25.30(0.05) & \textbf{24.07(0.16)}&\textbf{0.82(0.03)}&\textbf{0.79(0.03)}     \\ \hline
$(.5,.5)$-GAN+ $\ell_1$   & 3.01\text{e-}3(2.81\text{e-}5) & \textbf{ 3.98\text{e-}3(4.60\text{e-}5)}& 25.27(0.04) & \textbf{24.07(0.05)}&0.79(0.04)&\textbf{0.80(0.03)} \\ \hline
\end{tabular}

\label{tbl: sim_denoise}
\end{table}

In addition, Appendix ``\nameref{sec:cluster}" also shows an example that GAN with $\ell_1$-Autoencoder helps heterogeneous conformation clustering. 

\subsubsection{\textbf{Robustness under contamination}}
 \label{sec:robust}
We also consider the contamination model $\epsilon\neq 0$ and $Q$ from purely noisy images. We randomly replace partial samples of our training dataset of RNAP by noise to test whether our model is robust or not. There are three types to test: (A) Only replacing the clean reference images. It implies the reference images are wrong or missing, such that we do not have the reference images to compare. This is the worst contamination case. (B) Only replacing the noisy images. It means the Cryo-EM images which the machine produces are broken. (C) Replacing both, which indicates both A and B happen. The latter two are mild contamination cases, especially C that replaces both reference and noisy images by Gaussian noise whose $\ell_1$ or $\ell_2$ loss is thus well-controlled. 

Here we test our robustness of various deep learning based methods using the RNAP data of SNR 0.1, and the former \HLL{three types of contamination} is applied to randomly replace the samples in the proportion of $\epsilon\in \{0.1, 0.2, 0.3\}$ of the whole dataset. 

Fig. \ref{fig4}(e), (f) and (g) compare the robustness of different methods. In all the cases, some $\beta$-GANs ($(.5,.5)$- and $(1,1)$-) with $\ell_1$-Autoencoder exhibit relatively universal robustness. Particularly, (1) The MSE with $\ell_1$ loss is less than the MSE with $\ell_2$ loss, which represents the $\ell_1$ loss is more robust than $\ell_2$ as desired. (2) The Autoencoder method in $\ell_2$ loss and WGANgp show certain robustness in case B and C but are largely influenced by contamination in case A (shown in Figure \ref{fig4} (e)), indicating the most serious damage arising from type A, merely replacing only the reference image by Gaussian noise. The reason is that the $\ell_2$ Autoencoder and WGANgp method are confused by the wrong reference images so that they can not learn the mapping from data distribution to reference distribution accurately. (3) In the type C, the standard deviations of the five best models are larger compared the other two types. The contamination of both noisy $y$ and clean $x$ images influence the the stability of model more than the other two types. 

Furthermore, we take an example in type A contamination with $\epsilon=0.1$ for 3D reconstruction. The 3D reconstruction in denoised images with $(.5,.5)$-GAN + $\ell_1$ and $l_2$-Autoencoder are shown in Fig.~\ref{fig4}(h) and (j), and related FSC curve is \HLL{Fig.~\ref{fig4}(i)}. Specifically, on the one hand, the blue FSC curve of $\ell_2$-Autoencoder doesn't drop, which leads to the worse reconstruction; on the other hand, the red FSC curve of $(.5,.5)$-GAN + $\ell_1$ drops quickly but begins to rise again, whose reason is that some unclear detail of structure mixed angular information in reconstruction. When applying 0.143 cutoff criterion (dashed line in FSC curve), the resolution of $(.5,.5)$-GAN + $\ell_1$ is about 4$\mathring{A}$. Although reconstruction of images and final resolution is not better than the clean images, it is much clearer than $\ell_2$-Autoencoder which totally fails in the contamination case. The outcome of the reconstruction demonstrates that $(.5,.5)$-GAN + $\ell_1$ is relatively robust, whose 3D result is consistent with the clean image reconstruction.

\HLL{In summary, some $(\alpha,\beta)$-GANs methods, such as the ($(.5,.5)$-GAN and $(1,1)$-GAN, with $\ell_1$-Autoencoder are more resistant to sample contamination, which are better to be applied into the denoising of Cryo-EM data.}

\subsection{\textbf{Results for EMPIAR-10028}} \label{sec:empire}

The following Fig. \ref{fig5}(a) and (b) show the denoising results by different deep learning methods in experimental data: $\ell_1$ or $\ell_2$ Autoencoders, JS-GAN ($(0,0)$-GAN), WGANgp, and $(\alpha,\beta)$-GAN, where we add $\ell_1$ loss in all of the GAN-based structures. Although the Autoencoder can grasp the shape of macromolecules, it is a little blur in some parts. \HLL{What is more, WGANgp and $(.5,.5)$-GAN perform better than other deep learning methods according to MSE and PSNR, which is largely consistent with the result of the RNAP dataset.} The improvements of such GANs over pure Autoencoders lie in their ability of utilizing structural information among similar images to learn the data distribution better.

Finally, \HLL{we implement reconstruction} via RELION of 100000 images, which denoised by $(.5,.5)$-GAN +$\ell_1$. \HLL{The parameters are the same as the ones set in the paper~(\cite{wong2014cryo}).} The reconstruction results are shown in Fig. \ref{fig5}(c). It is demonstrated that the final resolution is 3.20$\mathring{A}$, which is derived by FSC curve in Fig. \ref{fig5}(d) using the same 0.143 cutoff (dashed line) to choose the final resolution. We note that the final resolution by RELION after denoising is as good as the original resolution 3.20$\mathring{A}$ reported in~\cite{wong2014cryo}.



\begin{figure}[t]
\centering
\includegraphics[width=5.5in]{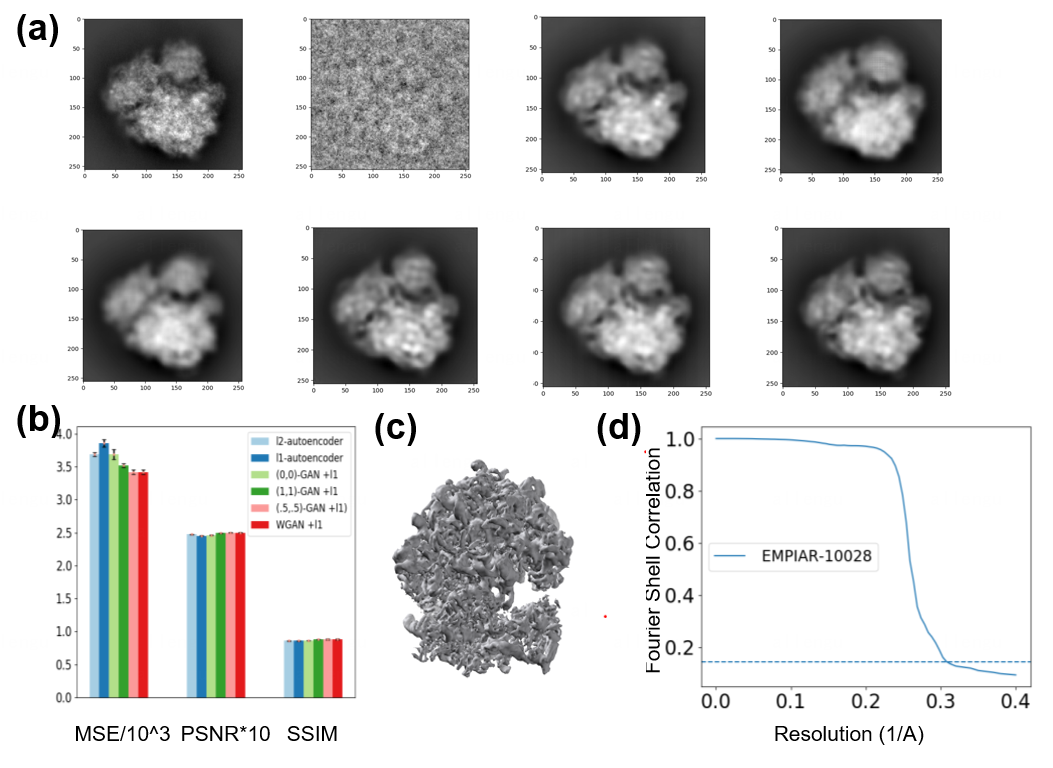}
\caption{Results for EMPIAR-10028. (a) Comparison in EMPIAR-10028 dataset in different deep learning methods (from left to right, top to bottom): clean image, noisy image, $\ell_1$-Autoencoder, $\ell_2$-Autoencoder, $(0,0)$-GAN + $\ell_1$, $(1,1)$-GAN + $\ell_1$, $(.5,.5)$-GAN + $\ell_1$,  WGANgp + $\ell_1$. (b) is the MSE, PSNR and SSIM in different denoised methods. (c) and (d) are the 3D-reconstruction of denoised images by $(.5,.5)$-GAN + $\ell_1$ and the FSC curve, respectively. The resolution of reconstruction from $(.5,.5)$-GAN + $\ell_1$ denoised images is 3.20$\mathring{A}$, which is as good as the original resolution.}
\label{fig5}
\end{figure}

\section{\textbf{Conclusion and Discussion}} \label{sec:conclusion}
In this chapter, we set a connection between the traditional image forward model and Huber contamination model in solving the complex contamination in the Cryo-EM dataset. \HLL{The joint training of Autoencoder and GAN has been proved to substantially improve the performance in Cryo-EM image denoising. In this joint training scheme, the reconstruction loss of Autoencoder helps GAN to avoid mode collapse and stabilize training. GAN further helps Autoencoder in denoising by utilizing the highly correlated Cryo-EM images since they are 2D projections of one or a few 3D molecular conformations}. To overcome the low Signal-to-Noise Ratio challenge in Cryo-EM images, joint training of $\ell_1$-Autoencoder combined with $(.5,.5)$-GAN, $(1,1)$-GAN, and WGAN with gradient penalty is often among the best performance in terms of MSE, PSNR, and SSIM when the data is contamination-free. However, when a portion of data is contaminated, especially when the reference data is contaminated, WGAN with $\ell_1$-Autoencoder may suffer from the significant deterioration of reconstruction accuracy. Therefore, robust $\ell_1$ Autoencoder combined with robust GANs ($(.5,.5)$-GAN and $(1,1)$-GAN) are the overall best choices for robust denoising with contaminated and high noise datasets.  

\HLL{Part of the results in this chapter is based on a technical report (\cite{gu20gan}). Most of the deep learning-based techniques in image denoising need reference data, limiting themselves in the application of Cryo-EM denoising.} For example, in our experimental dataset EMPIAR-10028, the reference data is generated by the cryoSPARC, which itself becomes problematic in highly heterogeneous conformations. Therefore, the reference image we learn may follow a fake distribution. How to denoise without the reference image thus becomes a significant problem. It is still open how to adapt to different experiments and those without reference images. In order to overcome this drawback, an idea called ``image-blind denoising" was offered by the literature\HLL{~(\cite{lehtinen2018noise2noise,krull2019noise2void})}, which viewed the noisy image or void image as the reference image to denoise. Besides,~\cite{chen2018image} tried to extract the noise distribution from the noisy image and gain denoised images through removing the noise for noisy data;~\cite{quan2020self2self} augmented the data by Bernoulli sampling and denoise image with dropout. Nevertheless, all of the methods need noise is independent of the elements themselves. Thus it is hard to remove noise in Cryo-EM because the noise from ice and machine is related to the particles.

In addition, for reconstruction problems in Cryo-EM,~\cite{zhong2020reconstructing} \HLL{proposed an end-to-end} 3D reconstruction approach based on the network from Cryo-EM images, where they attempt to borrow the Variational Autoencoder (VAE) to approximate the forward reconstruction model and recover the 3D structure directly by combining the angle information and image information learned from data. This is one future direction to pursue. 



\section{\textbf{Appendix}}\label{sec:appendix}

\subsection*{\textbf{Influence of parameter($\alpha, \beta$) brings in $\beta$-GAN}} \label{sec:betagans}
\HLL{In this part,} we have applied $\beta$-GAN into denoising problem. How to pick up a good parameter: $(\alpha, \beta)$ in the $\beta$-GAN becomes an important issue. Therefore, \HLL{we investigate} the impact of the parameter ($\alpha, \beta$) on the outcome of denoising.  We choose eight significant groups of $\alpha, \beta$. Our result is shown in Table \ref{tbl: alpha-beta-result}. It is demonstrated that the effect of these groups in different parameters is not large. The best result appears in $\alpha= 1, \beta =1$ and $\alpha= 0.5, \beta =0.5$ 

\begin{table}[H]\scriptsize
\renewcommand\arraystretch{1.5}
\centering
\caption{The result of $\beta$-GANs with ResNet architecture: MSE, PSNR and SSIM  of different $(\alpha, \beta)$ in $\beta$-GAN under various levels of Gaussian noise corruption in RNAP dataset.}
\begin{tabular}{|c|c|c|c|c|c|c|}
\hline
 & \multicolumn{2}{c|}{MSE} & \multicolumn{2}{c|}{PSNR} & \multicolumn{2}{c|}{SSIM} \\ \hline
  Parameter/SNR &  0.1 &0.05  &0.1 & 0.05 &0.1 & 0.05 \\   
\hline
$\alpha =1, \beta=1$  &\textbf{2.99\text{e-}3(3.51\text{e-}5)} & 4.01\text{e-}3(1.54\text{e-}4) &  \textbf{25.30(0.05) }& \textbf{24.07(0.16)}&\textbf{0.82(0.03)}&0.79(0.03)     \\ \hline
$\alpha =0.5, \beta=0.5$ & 3.01\text{e-}3(2.81\text{e-}5) & \textbf{ 3.98\text{e-}3(4.60\text{e-}5)}& 25.27(0.04) & \textbf{24.07(0.05)}&0.79(0.04)&\textbf{0.80(0.03)} \\ \hline
$\alpha =-0.5, \beta=-0.5$  & 3.02\text{e-}3(1.69\text{e-}5) & 4.15\text{e-}3(5.05\text{e-}5)& 25.27(0.02)& 23.91(0.05) &0.80(0.03)&\textbf{0.80(0.03)} \\ \hline
$\alpha =-1, \beta=-1$ & 3.05\text{e-}3(3.54\text{e-}5) & 4.12\text{e-}3(8.30\text{e-}5) & 25.23(0.05) & 23.93(0.08)   &0.80(0.05)&0.77(0.04)  \\ \hline
$\alpha =1, \beta=-1$& 3.05\text{e-}3(4.30\text{e-}5) & 4.10\text{e-}3(5.80\text{e-}5) & 25.24(0.06) & 23.96(0.06) &\textbf{0.82(0.02)}&0.76(0.03) \\ \hline
$\alpha =0.5, \beta=-0.5$& 3.09\text{e-}3(6.79\text{e-}5) &  4.05\text{e-}3(6.10\text{e-}5)& 25.17(0.04) & 24.01(0.06)&0.79(0.04)&0.77(0.05) \\ \hline
$\alpha =0, \beta=0$   & 3.06\text{e-}3(5.76\text{e-}5) & 4.02\text{e-}3(5.67\text{e-}4) & 25.23(0.04) & 24.00(0.06)&0.78(0.03)&0.78(0.03)     \\ \hline
$\alpha =0.1, \beta=-0.1$  & 3.07\text{e-}3(5.62\text{e-}5) & 4.05\text{e-}3(8.55\text{e-}5)& 25.23(0.08) & 23.98(0.04) &0.78(0.02)&0.79(0.03) \\ \hline

\end{tabular}

\label{tbl: alpha-beta-result}
\end{table}

\subsection*{\textbf{Clustering to solve the conformational heterogeneity}} \label{sec:cluster}
In this part, we try to analyze whether the denoised result is good in solving conformation heterogeneity in simulated RNAP dataset. Specifically, for heterogeneous conformations in simulation data,  we mainly choose the following two typical conformations: \emph{open} and \emph{close} conformations \HLL{(the leftmost and rightmost conformations in Fig. \ref{fig:5conf})} as our testing data. Our goal is to distinguish these two classes of conformations. However, different from the paper\HLL{~(\cite{xian2018data})}, we do not have the template images to calculate the distance matrix, so what we try is unsupervised learning -- clustering. Our clustering method is firstly using manifold learning: Isomap (\cite{tenenbaum2000global}) to reduce the dimension of the denoised images, then make use of $k$-Means ($k=2$) to group the different conformations. 

\begin{figure}[htbp]
\centering
\includegraphics[width=6in]{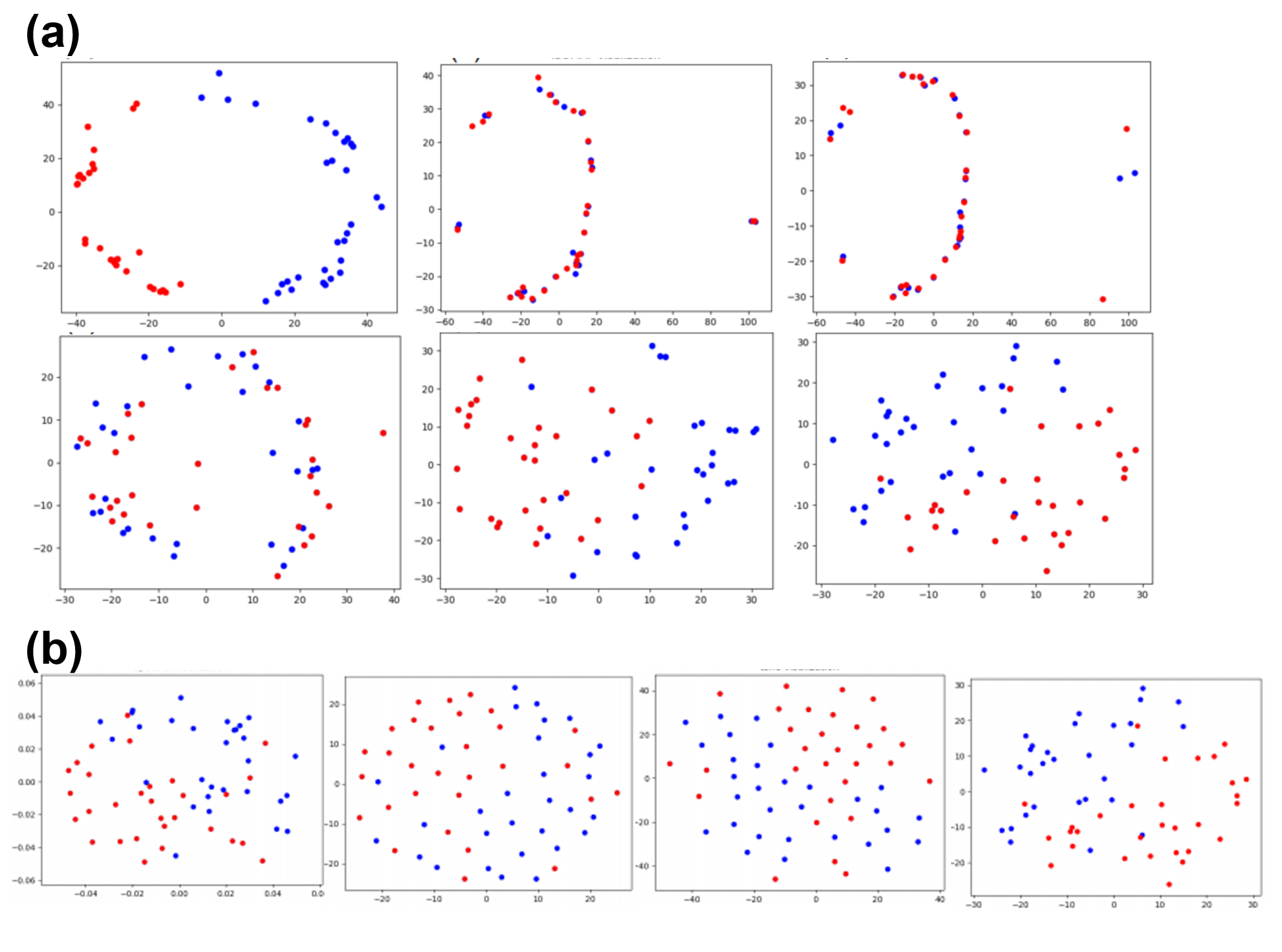}
\caption{ 2D visualization of 2-conforamtional images in manifold learning. Red point, blue point separately represent the open and closed conformation. (a) is 2D visualization of 2-conformation image by ISOMAP in different methods (from the left and top to the right and bottom): clean image, BM3D, KSVD, $\ell_2$-Autoencoder, $(1,1)$-GAN+ $\ell_1$, WGANgp+ $\ell_1$. (b) is 2D visualization of 2-conformation image in different manifold learning methods (from left to right): Spectral methods, MDS, TSNE, and ISOMAP.}
\label{fig: dif_clustering}
\end{figure}

The Fig.~\ref{fig: dif_clustering}(a) displays the 2D visualizations of two conformations about the clustering effect in different denoised methods. \HLL{Here the SNR of noisy data is 0.05.} In correspondence to  those visualizations, the accuracy of competitive methods is reported: $(1,1)$-GAN$+\ell_1$: $54/60$ (54 clustering correctly in 60), WGANgp$+\ell_1$: $54/60$, $\ell_2$-Autoencoder: $44/60$, BM3D: $34/60$, and KSVD: $36/60$. This result shows that: clean images separate well; $(\alpha,\beta)$-GAN and WGANgp with $l_1$ Autoencoder can distinguish the open and close structure partially, although there exists several wrong points; $\ell_2$-Autoencoder and traditional techniques have poor performance because it is hard to detect the clamp shape.

Furthermore, the reason we use Isomap is it performs the best in our case and comparisons of different manifold learning methods are shown in Fig.~\ref{fig: dif_clustering}(b). It demonstrates that blue and red points separate most in the graph of ISOMAP. Specifically, the accuracy of these four methods are $50/60$ (spectral method), $46/50$ (MDS), $46/50$ (TSNE), and  $54/60$ (ISOMAP). it shown that Isomap can distinguish best in the two structures' images compared to other methods: such as the Spectral method (\cite{ng2002spectral}), MDS (\cite{cox2008multidimensional}), and TSNE (\cite{maaten2008visualizing}). 

\subsection*{\textbf{Convolution network}} \label{sec:convnet}
\begin{figure}[t] 
\centering
\includegraphics[width=6in]{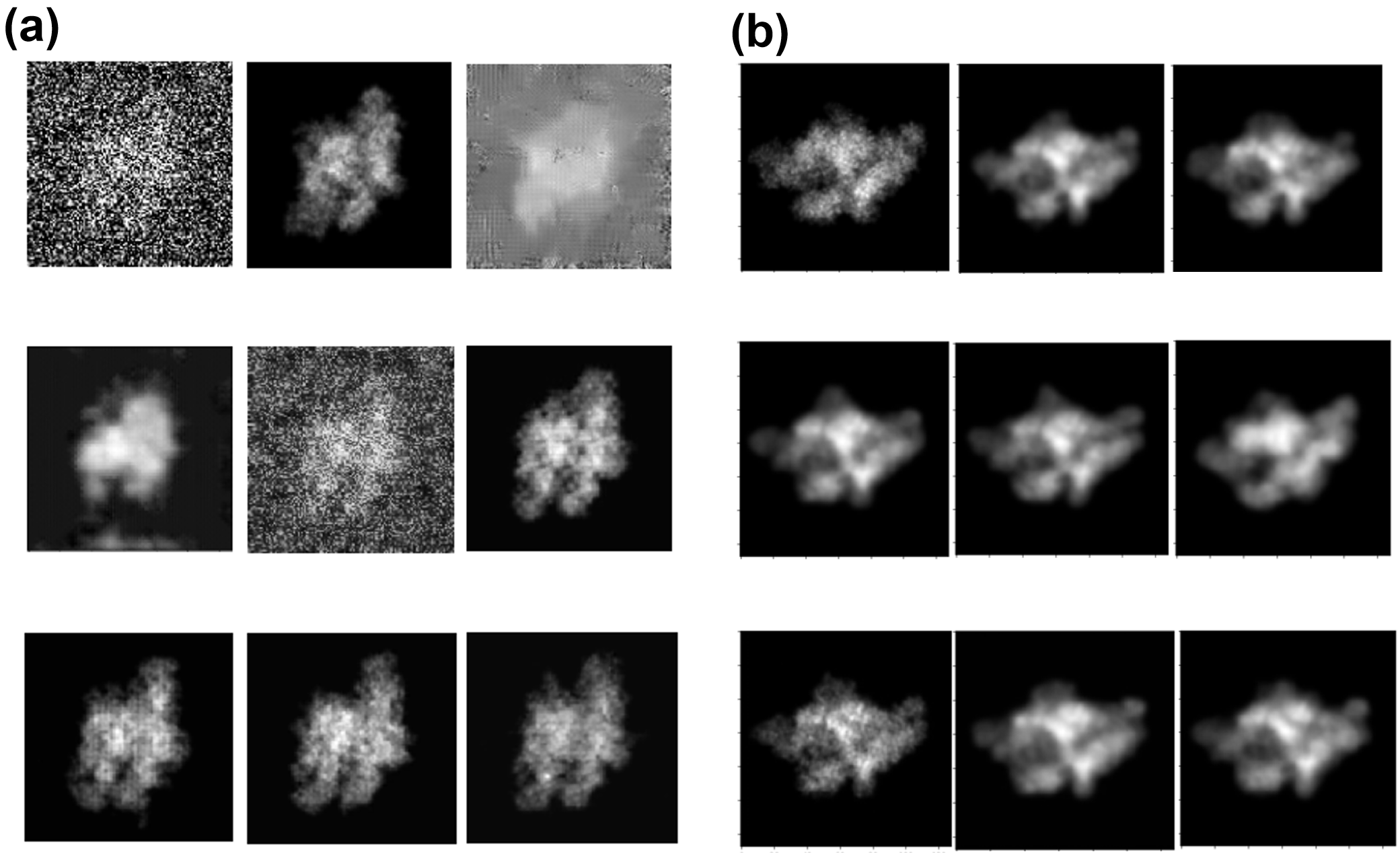}
\caption{(a) Denoised images with convolution network without ResNet structure in different methods in RNAP dataset with SNR 0.05 (from left to right, top to bottom): clean, noisy, BM3D, $\ell_2$-Autoencoder, KSVD, JS-GAN + $\ell_1$, WGANgp + $\ell_1$, $(1,-1)$-GAN + $\ell_1$, $(.5, -.5)$-GAN + $\ell_1$. (b) Denoised and reference images in different regularization $\lambda$ (we use $(.5,.5)$-GAN +$\lambda$ $\ell_1$ as an example) in correpsonding to Table \ref{tab:lambda}. From left to right, top to bottom, the image is: Clean image, $\lambda=0.1$, $\lambda=1$, $\lambda=5$, $\lambda=10$, $\lambda=50$, $\lambda=100$, $\lambda=500$ , $\lambda=10000$}
\label{fig:dcgan}
\end{figure}

We present the result of simple deep convolution network (remove the ResNet block), the performances in all of criterion are worse than performances of the residue's architecture work. Table \ref{tab:conv} compares the MSE and PSNR performance of various methods in the RNAP dataset with SNR $0.1$ and $0.05$. And Fig. \ref{fig:dcgan}(a) displays the denoised image of different methods in the RNAP dataset with SNR $0.05$. It shows the advantage of residue structure in our GAN-based denoising Cryo-EM problem.

\begin{table}[H] \footnotesize
\renewcommand\arraystretch{1.5}
\centering
\caption{\label{tab:conv}MSE and PSNR of different models under various levels of Gaussian noise corruption in RNAP dataset, where the architecture of GANs or Autoencoders are simply convolution network.} 
\begin{tabular}{|c|c|c|c|c|}
\hline
 & \multicolumn{2}{c|}{MSE} & \multicolumn{2}{c|}{PSNR}  \\ \hline
  Method/SNR &  0.1 &0.05  &0.1 & 0.05 \\   
\hline
BM3D  & 3.5\text{e-}2(7.8\text{e-}3)& 5.9\text{e-}2(9.9\text{e-}3) & 14.535(0.1452) & 12.134(0.1369)     \\ \hline
KSVD  & 1.8\text{e-}2(6.6\text{e-}3) & 3.5\text{e-}2(7.6\text{e-}3)& 17.570(0.1578)& 14.609(0.1414)  \\ \hline
Non-local means & 5.0\text{e-}2(5.5\text{e-}3)&5.8\text{e-}2(8.9\text{e-}3)&13.040(0.4935)&12.404(0.6498)  \\ \hline
CWF  & 2.5\text{e-}2(2.0\text{e-}3) & 9.3\text{e-}3(8.8\text{e-}4)& 16.059(0.3253)& 20.314(0.4129)  \\ \hline
$\ell_2$-Autoencoder  & 4.0\text{e-}3(6.0\text{e-}4) & 6.7\text{e-}3(9.0\text{e-}4)& 24.202(0.6414)& 21.739(0.7219) \\ \hline
$(0,0)$-GAN +$\ell_1$  & 3.8\text{e-}3(6.0\text{e-}4) & 5.6\text{e-}3(8.0\text{e-}4) & 24.265(0.6537) & 22.594(0.6314)     \\ \hline
WGANgp+$\ell_1$& \textbf{3.1\text{e-}3(5.0\text{e-}4)} & 5.0\text{e-}3(8.0\text{e-}4) & \textbf{25.086(0.6458)} & 23.010(0.6977)  \\ \hline
$(1,-1)$-GAN +$\ell_1$& 3.4\text{e-}3(5.0\text{e-}4) & \textbf{4.9\text{e-}3(9.0\text{e-}4)} &  24.748(0.7233) & \textbf{23.116(0.7399)}     \\ \hline
$(.5,-.5)$-GAN +$\ell_1$ & 3.5\text{e-}3(5.0\text{e-}4) & 5.6\text{e-}3(9.0\text{e-}4)& 24.556(0.6272) & 22.575(0.6441) \\ \hline
\end{tabular}

\label{tbl_result}
\end{table}

\subsection*{\textbf{Test RNAP dataset with PGGAN strategy}}
\label{sec:PGGAN}
\HLL{PGGAN~(\cite{karras2018progressive}) is a popular method to generate high-resolution images from low resolution ones by gradually adding layers of generator and discriminator. It accelerates and stabilizes the model training. Since Cryo-EM images are in large pixel size that fits well the PGGAN method, here we choose its structure\footnote{We set the same architecture and parameters as \url{https://github.com/nashory/pggan-pytorch} and the input image size is $128 \times 128$.} instead of the ResNet and convolution structures above to denoise Cryo-EM images.} Our experiments partially demonstrate two things: 1) the denoised images sharpen more, though the MSE changes to be higher. 2) we do not need to add $\ell_1$ regularization to make model training stable; it can also detect the outlier of images for both real and simulated data without regularization. 

\begin{figure}[htbp]
\centering

\includegraphics[width=6in]{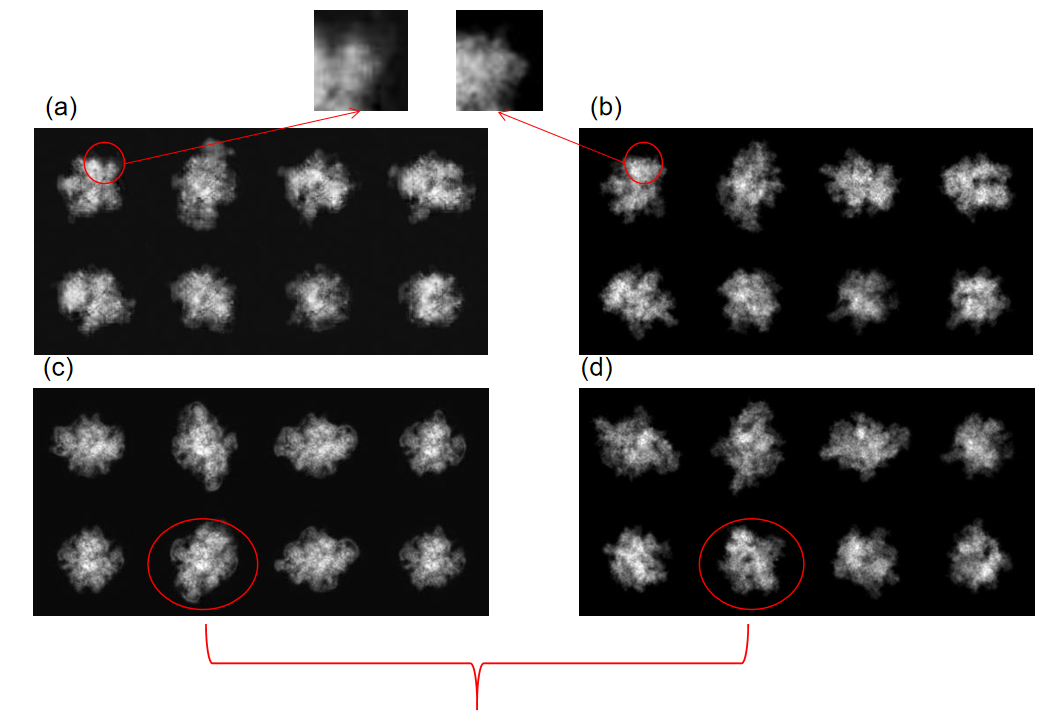}

\caption{\HLL{Denoised and reference images by PGGAN instead of simple ResNet and Convolution structure in RNAP dataset with SNR 0.05. The PGGAN strategy is tested in two objective functions: WGANgp + $\ell_1$ and WGANgp. (a) and (b) are denoised and reference images using PGGAN with WGANgp + $\ell_1$; (c) and (d) are denoised and reference images using PGGAN in WGANgp, respectively. Specifically, the images highlighted in red color show the structural difference between denoised images and reference images. It demonstrates that denoised images are different from reference images using PGGAN strategy.}}
\label{fig:pggan}
\end{figure}

\HLL{In detail, based on the PGGAN architecture and parameters, we test the following two objective functions developed in the section ``\nameref{sec:noisy modeling}": WGANgp and WGANgp + $\ell_1$, in the RNAP simulated dataset with SNR $0.05$ as an example to explain.} The denoised images are presented in Fig. \ref{fig:pggan}; it is noted that the model is hard to collapse regardless of adding $\ell_1$ regularization. The MSE of adding regularization is $8.09\text{e-}3(1.46\text{e-}3)$, which is less than $1.01\text{e-}2(1.81\text{e-}3)$ without adding regularization. Nevertheless, both of them don't exceed the results based on the ResNet structure above. This shows that PGGAN architecture does not have more power than the ResNet structure. But an advantage of PGGAN lies in its efficiency in training. So it is an interesting problem to improve PGGAN toward the accuracy of ResNet structure.

Another thing that needs to highlight is MSE may not be a good criterion because denoised images by PGGAN are clearer in some details than the front methods we propose. This phenomenon is also shown in Appendix ``\nameref{sec:lambda}". So how to find a better criterion to evaluate the model and combine two strengths of ResNet-GAN and PGGAN await us to explore.

\subsection*{\textbf{Influence of the regularization parameter: $\lambda$}} \label{sec:lambda}
In this chapter, we add $\ell_1$ regularization to make model stable, but how to choose $\lambda$ of $\ell_1$ regularization becomes a significant problem. Here we take $(.5, .5)$-GAN to denosie in RNAP dataset with SNR $0.1$. According to some results in different $\lambda$ in Table \ref{tbl: lambda}, we find as the $\lambda$ tends to infinity, the MSE results tends to $\ell_1$-Autoencoder, which is reasonable. Also, the MSE result becomes the smallest as the $\lambda=10$.

\HLL{What’s more, an interesting phenomenon is found that a much clearer result could be obtained at $\lambda=100$ than that at $\lambda=10$, although the MSE is not the best (shown in the Fig. \ref{fig:dcgan}(b)).}
\begin{table}[H]
\centering
\caption{MSE, PSNR and SSIM of different $\lambda$ in (.5,.5)-GAN + $\lambda l_1$ in RNAP dataset. \label{tab:lambda}}
\begin{tabular}{|c|c|c|c|}
\hline
$\lambda$/criterion& MSE& PSNR&SSIM \\ \hline
0.1 & 3.06\text{e-}3(4.50\text{e-}5) & 25.22(0.07)& \textbf{0.82(0.06)}\\ \hline
1  & 3.05\text{e-}3(4.49\text{e-}5) & 25.24(0.06)& 0.81(0.05)\\ \hline
5  & 3.03\text{e-}3(2.80\text{e-}5) & 25.26(0.04) & 0.80(0.04)    \\ \hline
10 & \textbf{3.01\text{e-}3(2.81\text{e-}5)} & \textbf{25.27(0.04)} & 0.79(0.04)      \\ \hline
50 & 3.07\text{e-}3(3.95\text{e-}5) & 25.20(0.06)& 0.79(0.02) \\ \hline
100 & 3.11\text{e-}3(5.96\text{e-}5) &25.15(0.06) & 0.80(0.02)  \\ \hline
500 & 3.17\text{e-}3(5.83\text{e-}5) &25.01(0.07)&0.78(0.04) \\ \hline
10000 & 3.17\text{e-}3(2.90\text{e-}5) &25.03(0.04) &0.79(0.04) \\ \hline
\end{tabular}
\label{tbl: lambda}
\end{table}

\label{app:robusttable}


%
%




\addcontentsline{toc}{section}{\textbf{Reference}}

\bibliographystyle{spbasic} 
\bibliography{sample.bib}

\end{document}